# A No-Reference Medical Image Quality Assessment Method Based on Automated Distortion Recognition Technology: Application to Preprocessing in MRI-guided Radiotherapy

**Authors：**

Zilin Wang[1]

Shengqi Chen[1]

Jianrong Dai[2]

Shirui Qin[2]

Ying Cao[2]

Ruiao Zhao[2]

Huaguo Wu[1](Corresponding author, E-mail: wuguohua@bupt.edu.cn )

Yuan Tang[2] (Corresponding author, E-mail: tangyuan82@126.com )

Jiayun Chen[2] (Corresponding author, E-mail: grace_chenjy@163.com )

*These authors contributed equally to this work. Guohua Wu，Yuan Tang and Jiayun Chen, contributed equally as corresponding authors

[1]School of Electronic Engineering, Beijing University of Posts and Telecommunications, Beijing, China

[2]Department of Radiation Oncology, National Cancer Center/National Clinical Research Center for Cancer/Cancer Hospital, Chinese Academy of Medical Sciences and Peking Union Medical College, Beijing, China.

【Abstract】

**Objective:** To develop a no-reference image quality assessment method using automated distortion recognition to boost MRI-guided radiotherapy precision.

**Methods:** We analyzed 106,000 MR images from 10 patients with liver metastasis,

captured with the Elekta Unity MR-LINAC. Our No-Reference Quality Assessment Model includes: 1) image preprocessing to enhance visibility of key diagnostic features; 2) feature extraction and directional analysis using MSCN coefficients across four directions to capture textural attributes and gradients, vital for identifying image features and potential distortions; 3) integrative Quality Index (QI) calculation, which integrates features via AGGD parameter estimation and K-means clustering. The QI, based on a weighted MAD computation of directional scores, provides a comprehensive image quality measure, robust against outliers. Leave-One-Out Cross-Validation (LOO-CV) assessed model generalizability and performance. Tumor tracking algorithm performance was compared with and without preprocessing to verify tracking accuracy enhancements.

**Results:** Preprocessing significantly improved image quality, with the QI showing substantial positive changes and surpassing other metrics. After normalization, the QI's average value was 79.6 times higher than CNR, indicating improved image definition and contrast. It also showed higher sensitivity in detail recognition with average values 6.5 times and 1.7 times higher than Tenengrad gradient and entropy. The tumor tracking algorithm confirmed significant tracking accuracy improvements with preprocessed images, validating preprocessing effectiveness.

**Conclusions:** This study introduces a novel no-reference image quality evaluation method based on automated distortion recognition, offering a new quality control tool for MRIgRT tumor tracking. It enhances clinical application accuracy and facilitates medical image quality assessment standardization, with significant clinical and research value.

# 1. Introduction

Medical imaging is a cornerstone of modern healthcare, playing a vital role in diagnosis and treatment planning. Technologies such as X-ray, CT, PET, ultrasound, and MRI provide physicians with detailed insights into anatomical structures and pathological conditions. However, image quality can vary due to differences in equipment, scanning parameters, and operator expertise. For instance, MRI is susceptible to artifacts caused by mechanical vibrations, magnetic field inhomogeneities, and patient movement, all of which can compromise image integrity and clinical utility.

The integration of MRI with radiation therapy in MR-Linac systems introduces further complexities. Real-time imaging during treatment must account for physiological motions, such as breathing and heartbeats, that can introduce motion artifacts and blur tumor boundaries, complicating the therapeutic process[1]. Additionally, the need for rapid scanning to reduce treatment durations can result in noisy images that obscure diagnostic details.

Preprocessing of medical images is critical for enhancing image quality and interpretability. However, this process can inadvertently introduce distortions, affecting the accuracy of subsequent diagnostic and therapeutic applications. The impact of preprocessing on medical image quality highlights the urgent need for stringent quality assessment protocols and preprocessing validation mechanisms to ensure the reliability and consistency of medical images[2].

In the realm of image quality assessment (IQA), methods are typically categorized as subjective or objective. Subjective assessment relies on human evaluators, while objective assessment employs mathematical models to quantify image quality. Objective assessment is further distinguished into full reference (FR), reduced reference (RR), and no reference (NR) methods[3-7]. Traditional IQA approaches often require a reference image for comparison, which is frequently impractical, particularly in dynamic or real-time imaging scenarios[8], this reality has heightened the importance of No-Reference Image Quality Assessment (NR-IQA)in medical imaging[9].

With the continuous advancement of computer vision and statistical analysis technologies, a series of mature metrics and methods for NR-IQA have been developed in the field of natural images. These methods effectively assess the types and degrees of distortions by leveraging statistical features and perceptual attributes of images. In recent years, research has gradually extended to the field of medical imaging, and to address the unique image quality needs in this domain, some improved NR-IQA methods specifically tailored for medical images have emerged. In this context, Kastryulin et al. (2022)[10] evaluated various IQA metrics and found that metrics such as DISTS, HaarPSI, VSI, and $FID_{VGG16}$ showed a strong correlation with radiologists' quality judgments, indicating high reliability in MRI image quality assessment. However, many existing studies in medical image quality still require the development of task-specific models for classification and estimation[11,12]. The methodologies put forward by Liu et al. (2020)[13], Kustner et al. (2018)[14], and Largent et al. (2021)[15] provide a comprehensive quality rating or score, whereas others, such as those by Kim et al. (2019)[16], Ye et al. (2014)[17], Xue et al. (2013)[18] and Li et al. (2017)[19], rely on ground truth comparisons or mean opinion scores (MOS). These latter approaches, while valuable, demand substantial human input and are subject to variability, which can be a significant drawback in clinical settings where consistency and efficiency are crucial.

Despite progress in the field, current NR-IQA methods face considerable hurdles in medical imaging. These methods are often tailored to specific tasks, narrowing their applicability and flexibility for various clinical uses. Typically intended for natural images, their efficacy is frequently diminished when transferred to medical imaging, which grapples with distinct distortions like noise, motion artifacts, and anisotropic distortions stemming from scanning directions or device parameters. Moreover, methods dependent on ground-truth comparisons or mean opinion scores (MOS) are variable and resource-intensive, largely due to the significant human input they require. These constraints highlight an urgent need for more robust, adaptable, and automated methods capable of providing dependable quality assessments in diverse medical imaging contexts, including the unique anisotropic distortions present in medical images. This is crucial to ensure that clinical decisions, such as tumor detection and treatment planning, rely on precise and high-quality imaging data.

In this work, we present an innovative reference-free medical image quality assessment method specifically designed to tackle the unique challenges posed by medical imaging. By leveraging Natural Scene Statistics (NSS) and drawing inspired by the principles of the Blind/Reference-free Image Spatial Quality Evaluator (BRISQUE)[20], our approach moves away from traditional distortion-specific feature calculations, such as those for artifacts[21,22], blurring[23,24]. Instead, it offers a comprehensive measure of "naturalness" degradation within medical images, which is vital for preserving diagnostic integrity. Utilizing locally normalized luminance coefficients, our method conducts an extensive statistical analysis of scene features, making it applicable to a broad spectrum of distortions commonly encountered in medical imaging. This versatility enables robust assessment across various clinical scenarios, ensuring that the method can reliably evaluate image quality within the complex and variable landscape of medical imaging. This approach is specially tailored to meet the unique demands of medical imaging, establishing a standardized and authoritative protocol for medical image preprocessing. It serves as a critical mechanism for validating the effectiveness of preprocessing techniques in specific tasks, such as tumor tracking or diagnostic evaluations in radiotherapy. By ensuring that these preprocessing methods adhere to stringent quality benchmarks, our method lays a solid foundation essential for the success of subsequent clinical applications, ultimately enhancing the precision and reliability of radiotherapy procedures.

## 2. Materials and Methods

### 2.1 Image Acquisition

Our study leveraged the Elekta Unity MR-Linac system, a cutting-edge device that integrates high-resolution MRI technology with a linear accelerator, to capture MRI images from patients receiving MRI-guided radiotherapy (MRIgRT). The Elekta Unity, featuring a 1.5 Tesla MRI and an advanced linear accelerator, transcends the constraints of conventional X-ray and CT imaging, especially in the delineation of soft tissues. Its real-time tracking and adaptive radiation therapy capabilities are pivotal for improving treatment accuracy and outcomes.

The dataset for this investigation was obtained using the Elekta Unity system, employing a high-resolution, rapid MRI technique that incorporates Beam Eye View (BEV) and Beam Path View (BPV) fusion data, named as 3.5 D MRI. This innovative imaging approach was designed for precise real-time tracking and dynamic radiation therapy adjustments, following established protocols from a previously published study[25]. The data encompassed images from ten patients with liver metastases, each image sequence consisting of 50 frames. The comprehensive dataset comprises 2,120 image sequences, totaling 106,000 MRI frames. Patients were categorized into two groups based on respiratory management strategies: one group used an abdominal compression belt to restrict abdominal movement (RAM group), while the other group breathed without restriction (FB group). The imaging technique provided high temporal and spatial resolution images, offering an invaluable dataset for the refinement and validation of tumor tracking methods in MRIgRT.

**2.2 Image Preprocessing**

Despite high image quality, some frames showed artifacts like variable brightness and noise, mandating preprocessing. Steps including grayscale normalization, gamma correction, and Gaussian filtering enhanced the dataset for accurate tumor tracking and analysis. These techniques were crucial for improving initial images' uneven brightness, poor contrast, and noise, ensuring the refined dataset met the standards necessary for precise tracking and treatment evaluation in tumor tracking research.

In this study, images underwent several specific preprocessing techniques to align with the requirements of the subsequent tumor tracking method[26], Grayscale normalization was applied to achieve consistent luminance, thereby ensuring reliable analysis. This was followed by gamma correction to enhance contrast, making tumors and tissues more discernible. Finally, Gaussian filtering was employed to reduce noise while preserving critical image features. The improved image quality significantly boosts the performance of the tracking algorithm, enabling faster and more precise tumor tracking.

**2.3 Distortion Recognition Techniques**

Distortion recognition techniques are pivotal for pinpointing and quantifying image artifacts that could impair diagnostic image quality[27]. They detect anomalies in image structure, brightness, and texture due to noise, motion, and equipment. A prevalent method involves statistical measures that flag deviations from typical image attributes[27]. Notably, the Mean Subtracted Contrast Normalized (MSCN) coefficient[28] has emerged as a reliable metric for assessing image quality and uncovering potential distortions.

The MSCN coefficient standardizes contrast and normalizes brightness variances, leading to a more uniform contrast level essential for detecting distortions and artifacts. It's calculated using the formula:

$$\hat{I}(i,j) = \frac{I(i,j) - \mu(i,j)}{\sigma(i,j) + C}, \quad (1)$$

where $i \in \{1,2,...,M\}$ and $j \in \{1,2,...,N\}$ are spatial indices, M and N are image dimensions. $C$ is a small constant to prevent division by zero, set as $C = 1$ in our application. The local mean $\mu(i,j)$ and local standard deviation $\sigma(i,j)$ are calculated as follows:

$$\mu(i,j) = \sum_{k=-K}^{K} \sum_{l=-L}^{L} w_{k,l} I(i+k, j+l), \quad (2)$$

$$\sigma(i,j) = \sqrt{\sum_{k=-K}^{K} \sum_{l=-L}^{L} w_{k,l} (I(i+k, j+l) - \mu(i,j))^2}. \quad (3)$$

In these equations, $w_{k,l}$ denotes a bidimensional, circularly symmetric Gaussian weighting function, confined within a window size of $(2K + 1) \times (2L + 1)(2K + 1) \times (2L + 1)$ and normalized to unit volume, with $K = L = 3$ in our implementation.

The MSCN coefficient is particularly advantageous in medical imaging, countering non-uniform brightness from varying scan parameters and patient movement. It enhances visibility of organ and tissue boundaries, crucial for diagnosis and treatment planning, and minimizes noise impact, improving the signal-to-noise ratio.

Figure 1 contrasts original and processed MRI images with their MSCN coefficient images, highlighting the coronal and sagittal views. The original images show brightness and contrast variations obscuring details, while the MSCN images reveal inconsistencies and noise. Preprocessing improves brightness and contrast uniformity, clarifying organ and tissue boundaries. After processing MSCN images display reduced noise and enhanced contrast, simplifying distortion and artifact identification. This comparison highlights the effectiveness of our preprocessing and the MSCN coefficient in enhancing medical image quality. By standardizing image quality and minimizing noise, our method supports accurate tumor tracking and reliable clinical outcomes.

While MSCN coefficients in undistorted natural images often follow a unit normal distribution, medical images may deviate due to smaller sample sizes or distortions[29]. We analyzed the Probability Density Function of MSCN coefficients (PDFMC) for both views before and after preprocessing. Figure 2 shows the MSCN coefficient distribution often deviates from Gaussian, indicating a need for adaptable methods to model its distribution in medical images. Preprocessing shifts PDFMC, increasing peak probability density and centralizing coefficients around zero, reflecting improved contrast normalization, reduced noise, and heightened image clarity—key for medical diagnostics. By understanding these nuances, we can develop more accurate MSCN coefficient distribution models, strengthening our no-reference image quality assessment methodology.

**2.4 No-Reference Quality Assessment Model**

In 2.3, the Mean Subtracted Contrast Normalized (MSCN) provides a global assessment of image properties, capturing overall variations in brightness and contrast. To gain a more detailed understanding of directional information within the image, we employed the method described in Figure 3. This approach involves multi-directional feature extraction and analysis, enabling us to capture structural information along different orientations, thereby providing a more nuanced evaluation of image quality.

**Step 1: Image Preprocessing**

In this step, we undertake preprocessing of the original image. The detailed processing particulars can be found in 2.2. These operations are intended to improve the clarity of the image, particularly highlighting diagnostic features such as tumor edges, to provide a superior basis for the subsequent quality assessment.

**Step 2: Feature Extraction and Directional Analysis**

**1）Directional Pixel Interaction Calculation：**

MR images exhibit directional structures with varying textural properties across different orientations, these features are vital for precise localization and analysis, particularly in tumor detection and diagnosis. To capture textural attributes across diverse directions, we calculate the product of adjacent pixel values on MSCN coefficients along four distinct directional vectors, as shown in Figure 4. Each direction in the figure represents a unique gradient and interaction type:

- **Central Pixel**: Represents the focal point of the neighborhood, denoted as $I(i,j)$, from which interactions with surrounding pixels are measured.

- **Right Diagonal ($D_{Right}$)**: Indicates the product with the pixel located one step to the right and one step down from the central pixel, as defined by:

$$D_{Right}(i,j) = \hat{I}(i,j)\hat{I}(i+1,j+1). \tag{4}$$

- **Horizontal ($H$)**: Reflects the product with the adjacent pixel to the right of the central pixel, expressed as:

$$H(i,j) = \hat{I}(i,j)\hat{I}(i,j+1). \tag{5}$$

- **Left Diagonal ($D_{Left}$)**: Corresponds to the product with the pixel situated one step to the left and one step above from the central pixel, given by:

$$D_{Left}(i,j) = \hat{I}(i,j)\hat{I}(i+1,j-1). \tag{6}$$

- **Vertical ($V$)**: Represents the interaction between the central pixel and the pixel located one step directly below it, calculated as:

$$V(i,j) = \hat{I}(i,j)\hat{I}(i+1,j). \tag{7}$$

Here, $i \in \{1,2,...,M\}$ and $j \in \{1,2,...,N\}$ represent the spatial indices within the image dimensions. This directional analysis captures different orientations' textural attributes and gradient variations, critical for identifying image features and potential distortions.

**2) AGGD Fitting and Parameter Estimation:**

The statistics of adjacent pixels products allow for the capture of unique image distortions. In order to tackle the challenge that the small sample scenarios in 2.3 do not satisfy the Gaussian distribution, we employ the Asymmetric Generalized Gaussian Distribution (AGGD)[30] for

statistical analysis. Assuming the AGGD with a zero mean, we model these pixel interactions:

$$f(x;\gamma,\beta_l,\beta_r) = \begin{cases} \dfrac{\gamma}{(\beta_l+\beta_r)\Gamma\left(\frac{1}{\gamma}\right)}\exp\left(-\left(\dfrac{-x}{\beta_l}\right)^{\gamma}\right) & \forall x \leq 0 \\ \dfrac{\gamma}{(\beta_l+\beta_r)\Gamma\left(\frac{1}{\gamma}\right)}\exp\left(-\left(\dfrac{x}{\beta_r}\right)^{\gamma}\right) & \forall x \geq 0 \end{cases}. \qquad (8)$$

From the AGGD fitting, we derive three key parameters: $\gamma$(sharpness), $\beta_l$(left scale), $\beta_r$(right scale) according to the formula:

$$\eta = (\beta_r - \beta_l)\dfrac{\Gamma\left(\frac{2}{\gamma}\right)}{\Gamma\left(\frac{1}{\gamma}\right)}. \qquad (9)$$

**3）K-means Clustering and Quality Metric Calculation：**

In medical image quality assessment, texture features at different angles and directions often reflect different quality of the image. To effectively integrate these features and address distortions arising from varying directions, we employed the K-means clustering algorithm[31], Using the moment matching method[32] to cluster the 12 sets of feature matrices obtained from AGGD fitting. Unlike conventional methods for natural images that directly analyze MSCN and AGGD features individually, our approach uses K-means clustering to categorize image quality features into three distinct classes: brightness consistency, structural integrity, and noise. This enables a more comprehensive representation of the anisotropic distortions inherent in medical images and their impact on overall image quality.

Each patient's image dataset includes numerous frames, each is a unique image entity. These frames, aligned with the body's condition, pathological status, and imaging parameters at the time of acquisition, constitute a set representing this patient's imaging session. For each frame, we compute 12 AGGD fitting parameters to capture variations in brightness and structural elements across different orientations. All feature vectors of this patient are compiled into a feature matrix and then subjected to K-means clustering. The AGGD parameters are Divided into K=3 clusters, each representing a distinct pattern of image quality features: brightness consistency, structural integrity, and noise. To characterize each cluster, we analyze the AGGD parameters at the cluster centers. For example, if a cluster center's shape parameter $\gamma$ significantly surpasses both the left and right scale parameters across all directions, that cluster is primarily driven by the shape parameter. This approach allows for meaningful categorization of image quality variations across frames.

When clusters display mixed features, we tackle this this complexity with a two-step approach: initial clustering of the entire dataset followed by a detailed analysis of clusters with mixed characteristics. After refinement, we establish initial weights based on the proportion of features within the hybrid cluster relative to others, leading to the final weight calculation based on the feature distribution within each cluster. We use iterative refinement to address any remaining confounding factors, ensuring that the final weights accurately represent the importance of various image features. This process enhances the accuracy and reliability of features extracting. It's important to note that the weight $w_1$ linked to the shape parameter $\gamma$ is positive, signifying its direct correlation with structural integrity—where higher $\gamma$ values generally denote more distinctly defined image structures. In contrast, the weights ($w_2$ and $w_3$) for the left scale parameter $\beta_l$ and the right scale parameter $\beta_r$ are negative. The cause of this situation lies in the fact that these parameters reflect the conditions of image distortion, such as uneven brightness and noise; thus, their lower values are associated with higher image quality. These weights capture the structural, brightness, and noise characteristics, forming the basis for subsequent quality assessment.

**Step 3: Integrative Image Quality Index Calculation**

In this phase, we outline the no-reference quality assessment model metrics. By using the weights from step 2 to weight the AGGD feature parameters $\gamma$, $\beta_l$, $\beta_r$, a quality score for each direction of the image is derived, as follows:

$$QI_{direction} = w_1 \times \gamma + w_2 \times \beta_l + w_3 \times \beta_r. \tag{10}$$

Then, we employ the Median Absolute Deviation (MAD) to integrate the quality scores across the different directional. The MAD is a robust statistical measure, adept at gauging data dispersion while significantly mitigating the disruptive influence of outliers[33]. The MAD is given by the formula:

$$MAD = median(|\, Qi - median(Q)\, |), \tag{11}$$

here $Qi$ signifies the quality score attributed to the image along a specific direction, with $median(Q)$ representing the median scores aggregated from all images within that directional assessment.

We proceed to synthesize the directional scores, utilizing MAD as a foundation for a weighted computation, thereby arriving at a comprehensive image quality score. The formula for this integrative score is:

$$QI = \frac{MAD_{horizontal}}{MAD_{total}} \times QI_{horizontal} + \frac{MAD_{vertical}}{MAD_{total}} \times QI_{vertical} + \frac{MAD_{diag_{left}}}{MAD_{total}} \times QI_{diag_{left}} + \frac{MAD_{diag_{right}}}{MAD_{total}} \times QI_{diag_{right}}, \tag{12}$$

where $MAD_{total} = MAD_{horizontal} + MAD_{vertical} + MAD_{diag_{left}} + MAD_{diag_{right}}$.

The resulting Quality Index $QI$ we obtain is a holistic measure of the image's overall quality, encompassing structural details, contrast, and distortions. Our empirical have consistently shown that this scoring methodology performs reliably across a broad spectrum of patient image datasets and various temporal snapshots, providing an accurate and dependable assessment of image quality. A significant advantage of this methodology is its reference-free nature, making it versatile and suitable for a wide range of scanning conditions and clinical scenarios. By incorporation of the MAD technique, we enhance the score's robustness, significantly reducing the impact of outliers. Meanwhile, the comprehensive QI score offers detailed insights into the specific quality characteristics of the image, thereby providing valuable support for clinical diagnosis and treatment decisions.

Our no-reference image quality assessment model underwent rigorous validation using an extensive dataset sourced from ten patients, encompassing a total of 106,000 MRI frames. To accurately assess the model's ability to generalize and its performance metrics, we employed a Leave-One-Out Cross-Validation (LOO-CV) strategy. The optimal weight parameters $w_1$, $w_2$ and $w_3$ for the model were determined based on the performance across all iterations, ensuring robust and generalized evaluation criteria.

## 3. Results

LOO-CV validated the generalizability and performance of the No-Reference Quality Assessment model. A detailed analysis of a single patient's dataset, consisting of 144 MRI sets with 50 frames each (totaling 7,200 frames), confirmed the efficacy of the optimal weight parameters. A meticulous comparison of image quality metrics before and after the preprocessing revealed significant insights. Figure 5 delineates the enhancement in image quality scores attributable to preprocessing. The left panel scrutinizes the quality scores of the initial 10 frames, whereas the right panel broadens the view to encompass scores from frames

11 to 50. Due to the instability at the beginning of the scan, the first 10 frames often exhibit fluctuations in image quality. We separated them to facilitate a clear observation of the effects on the Quality Index (QI) values. Even so, we observed that the QI values for both the first and last 10 frames were significantly higher after preprocessing compared to before preprocessing. A notable increase in quality scores is observed in after preprocessing images (Q2) compared to their before preprocessing counterparts (Q1), underscoring the substantial improvement in image quality achieved through our preprocessing strategies.

Upon closer inspection of Figure 5, it becomes clear that the initial frame of our dataset was intentionally given special consideration, with the first 10 frames undergoing a tailored analysis. This approach was adopted because the initial frames captured during the MRI acquisition process might not accurately reflect diagnostic quality due to potential activation inconsistencies. By deliberately excluding the first 10 frames from our comprehensive assessment, we can evaluate the quality of the subsequent frames with enhanced accuracy. Furthermore, juxtaposing the quality of many of the latter frames with that of the initial frame provides additional validation for the efficacy of the image quality scoring system presented in this study. This refined method guarantees more consistent and dependable outcomes within the quality assessment process, ultimately leading to a more robust analysis of image quality enhancements.

After confirming the image quality score, we proceeded to evaluate the impact of enhanced image quality on the performance of the automatic tracking algorithm. The experimental results clearly show that the after preprocessing images display significantly better tracking performance in automatic tracking algorithm compared to their before preprocessing counterparts, with substantial enhancements in both the stability and precision of the tracking[34].

In our analysis of image quality metrics, we compared several established methods, including Contrast-to-Noise Ratio (CNR)[35], Tenengrad gradient[36], Entropy[37], Naturalness Image Quality Evaluator (NIQE[38]), Perception-based Image Quality Evaluator (PIQE[39]), along with our innovative Quality Index (QI). Figure 6 provides a visual representation of the changes in these metrics following preprocessing. This comparison allows for a thorough understanding of how each metric responds to the quality improvements induced by our preprocessing steps.

To accurately portray the enhancements in image quality, specific guidelines were meticulously followed in the creation of Figure 6. For metrics where an increase in value denotes improvements (such as CNR, Tenengrad gradient, Entropy, and QI), the data were plotted directly, reflecting the actual increases. Conversely, for metrics like NIQE and PIQE, where a decrease in value signifies better quality, the data were inverted to present the negative values. This inversion was crucial to maintain a consistent interpretation of quality improvements across all metrics, allowing for a uniform representation of enhancements as upward shifts in the comparative analysis. This approach ensured a clear and coherent visualization, making it easier to discern the overall impact of preprocessing on image quality.

The analysis presented in Figure 6 underscores that the QI demonstrates the most significant positive change among all evaluated metrics, achieving a median value of 7.478 and a mean of 7.517. This improvement highlights the effectiveness of our proposed QI in accurately reflecting advancements in medical image quality, particularly in terms of diagnostic clarity. Additionally, the relatively low standard deviation of 0.875 further reinforces the consistency of these enhancements across different frames.

In contrast, conventional metrics exhibit limited substantial progress. The CNR records a median increase of 0.096 and an average of 0.093, suggesting it may not fully encompass the range of quality improvements critical for medical imaging. The Tenengrad gradient, which assesses image sharpness, reveals moderate enhancement with a median value of 1.004 and an average of 1.005. Meanwhile, the Entropy metric experiences a median increase of 2.678 and an average of 2.754, indicating an uptick in image complexity and detail, but this does not necessarily correlate to improved diagnostic quality. While these metrics provide insights into enhancements in image quality, our QI delivers a more comprehensive evaluation. With mean values that are 7.9 times, 6.5 times, and 1.7 times greater than those of CNR, Tenengrad, and Entropy respectively, the QI encompasses sharpness improvements as well as other critical facets of image quality, thus offering a more exhaustive and sensitive approach to assessing diagnostic image quality.

Significantly, metrics originally designed for assessing natural image quality, NIQE and PIQE, are exhibiting counterintuitive trends. NIQE reflects a mean shift of -0.926, while PIQE shows a more pronounced mean shift of -2.100. These results, which suggest a perceived decline rather than an improvement in image quality, are in stark contrast to the positive shifts recorded by metrics like QI. This inconsistency highlights potential limitations when employing NIQE and PIQE in medical imaging applications, where image characteristics and quality enhancement objectives can significantly differ from those found in natural image contexts. A thorough investigation of this unexpected finding will be addressed in the discussion section.

Our methodology augments the current frameworks of image quality evaluation in medical imaging, bolstering its practical clinical application. Given the unique challenges of medical image preprocessing, there is a need for metrics that accurately reflect quality improvements essential for diagnostic precision. Our approach addresses this need, offering a more refined and contextually relevant measure of image quality in medical imaging.

## 4. Discussion

The urgent need for reliable and precise methodologies for evaluating medical image quality is particularly vital in clinical environments, where accurate diagnoses and effective treatments rely on image clarity. Despite numerous quality assessment methods proposed in the medical imaging research domain, each possesses inherent limitations. This study introduces a comprehensive set of standardized methodologies for medical image processing, meticulously crafted to thoroughly assess the impact of preprocessing and guarantee the integrity of subsequent diagnostic and therapeutic interventions.

Our study's methodology surpasses traditional full-reference metrics like PSNR or SSIM, by forgoing the need for extensive datasets, aligning more closely with the practical constraints of clinical and medical research environments where high-quality reference images are often unattainable. Our method outperforms previous methods, including those by Liu et al. (2020)[13], Kustner et al. (2018)[14], and Largent et al. (2021)[15], by being finely attuned to the nuances of medical images. It emphasizes critical diagnostic elements, such as the accurate demarcation of tumor boundaries and the uniformity of tissue contrast, which are vital for precise clinical assessments.

Moreover, diverging from previous objective assessment models like BRISQUE[19,20], our method solely relies on natural scene statistics features extracted from undistorted medical images, sidelining the need for extensive subjective human perception scores. This advancement not only diminishes the resource demands of manpower and materials but also paves the way for a more objective and standardized approach to medical image quality assessment.

The acquisition of MR images is subject to a variety of influences, including scanner variability, scanning parameters, patient physiological conditions, and movement during scanning, which can lead to inconsistent image quality[40]. Our method effectively normalizes these variations by standardizing the image's structural, texture features, and intensity attributes, as illustrated in Figure 1. By integrating automation and standardized processing techniques, our approach accurately identifies and corrects a wide range of distortions and noises, ensuring a consistent level of image quality.

Furthermore, recognizing the privacy and ethical considerations associated with medical image data, as well as the diversity in equipment and protocols across medical institutions, our method avoids the need for large-scale, high-quality reference samples[7]. Instead, it provides a reliable and adaptable image quality assessment mechanism based on no-reference quality assessment technology, making it well-suited for various clinical environments. This innovative approach not only respects patient privacy and ethical standards but also accommodates the heterogeneity across different medical facilities, offering a versatile solution for image quality assessment in the clinical setting.

The Quality Index (QI) scores, as shown in Figure 5, rise significantly after preprocessing, indicating the method's sensitivity to quality improvements. With a mean difference of 7.517 and a low standard deviation of 0.875, the QI surpasses other metrics in reliability and effectiveness, as detailed in Figure 6. This makes it an ideal tool for enhancing clinical diagnosis and treatment planning, offering a nuanced measure of image quality improvements.

The differences observed in Figure 6 underscore the limitations of NIQE and PIQE when applied to medical imaging. These metrics are designed to capture "natural" aesthetics in images, often prioritizing global naturalness. This focus on natural image statistics causes NIQE and PIQE to overlook the contrast enhancement and local detail crucial for diagnosis, may interpret which intended to enhance diagnostic feature preprocessing techniques as degradations in image quality. By contrast, our QI is specifically adapted to the requirements of medical imaging, prioritizing local contrast and structural clarity essential for clinical evaluations. This approach ensures that the QI metric accurately reflects improvements in diagnostic relevance, making it a more suitable tool for assessing image quality in a medical context.

Our study highlights the adaptability and precision of automated processing techniques, presenting their adeptness at handling the diverse anatomical structures and pathological variations present in MR images. By dynamically adjusting processing strategies, our method ensures high sensitivity and consistent image quality evaluation, significantly enhancing the standardization of medical image quality assessment. This approach provides a robust foundation for subsequent diagnostic and therapeutic procedures. It guarantees the efficiency and reliability of medical images in clinical practice and strengthens the basis of clinical decision-making.

While this study presents significant findings, there is potential for further refinement and expansion. Future work should focus on broadening the dataset to include a wider range of clinical scenarios, enhancing the model's versatility and applicability. Ongoing optimization of the feature extraction and clustering algorithms will improve the model's computational efficiency and real-time capabilities, meeting the diverse demands of clinical applications. Additionally, exploring advanced multimodal image fusion techniques could increase the model's adaptability and robustness, especially in handling complex physiological conditions. Addressing these areas will be crucial for consolidating the model's clinical utility and advancing the field of medical image analysis.

## 5. Conclusion

Our study introduces a pioneering no-reference image quality assessment (NRIQA) method that enhances the precision and reliability of medical image evaluations, which is vital for applications such as tumor tracking. This method, bolstered by unsupervised learning and feature extraction techniques, has been validated for its robust performance across diverse patient datasets. It offers a standardized evaluation without the need for reference images and dynamically adapts to physiological changes during treatment, ensuring accurate image quality assessments crucial for clinical diagnosis and treatment planning. The study's findings underscore the transformative potential of this NR-IQA method in medical imaging, establishing a foundation for ongoing innovation in the field.

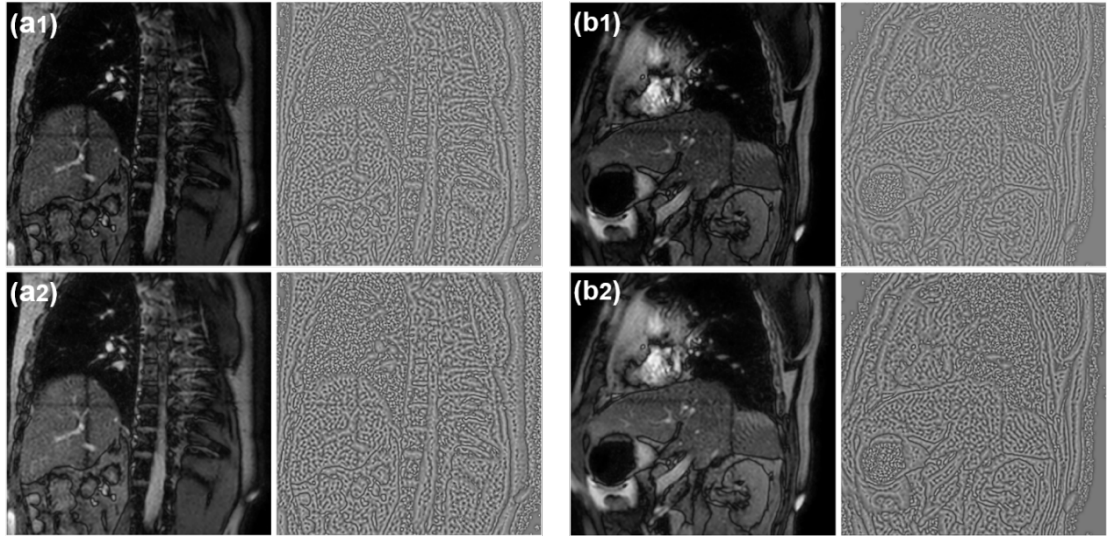

**Figure 1 Comparative Analysis of MRI Images in Coronal and Sagittal Views Before and After Preprocessing Along with Their MSCN Coefficient Images: (a1) Coronal view of the original MRI image and its corresponding MSCN visualization before preprocessing; (b1) Sagittal view of the original MRI image and its corresponding MSCN visualization before preprocessing; (a2) Coronal view of the MRI image and its MSCN visualization after preprocessing; (b2) Sagittal view of the MRI image and its MSCN visualization after preprocessing.**

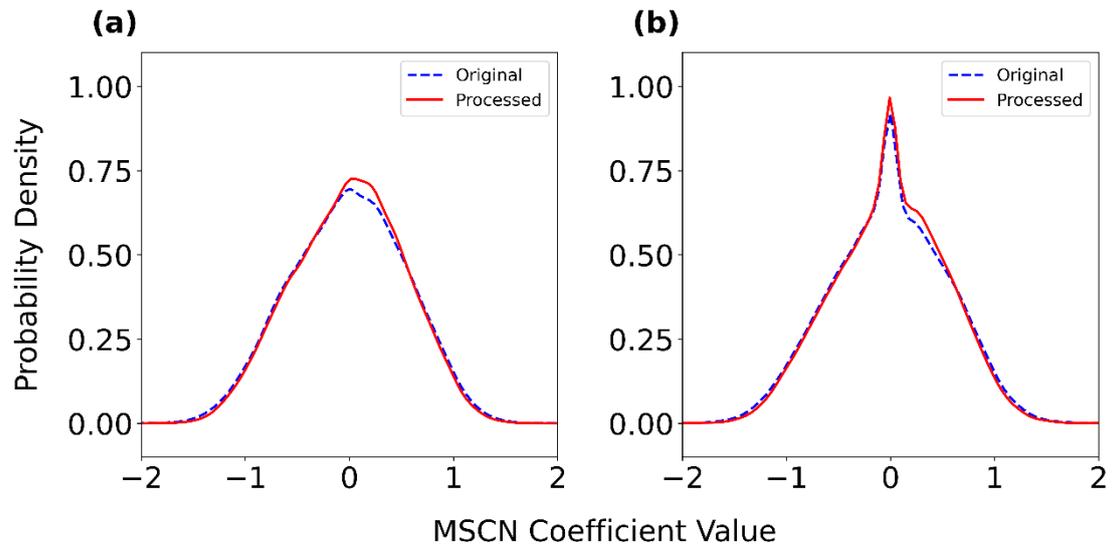

Figure 2 Probability Density Function of MSCN Coefficients (PDFMC) for Coronal and Sagittal Views. The left side represents the coronal view's PDFMC (a), and the right side corresponds to the sagittal view's PDFMC (b).

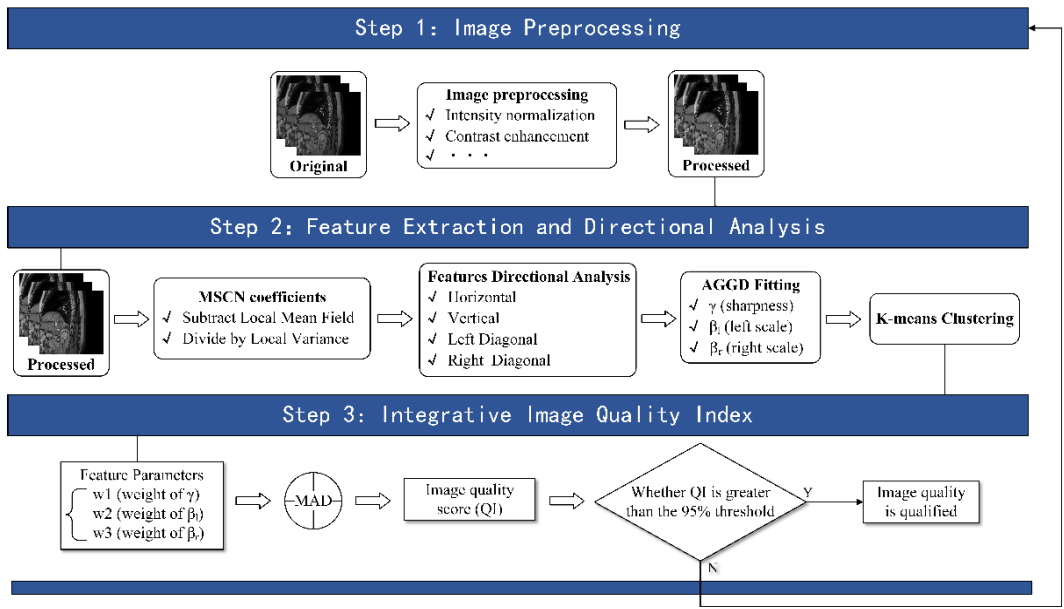

**Figure 3 The Workflow of the No-Reference Image Quality Assessment Method. Step 1: Image preprocessing to improve the visibility of key diagnostic features. Step 2: Feature extraction and directional analysis to evaluate image quality. Step 3: Calculate a integrative image quality score.**

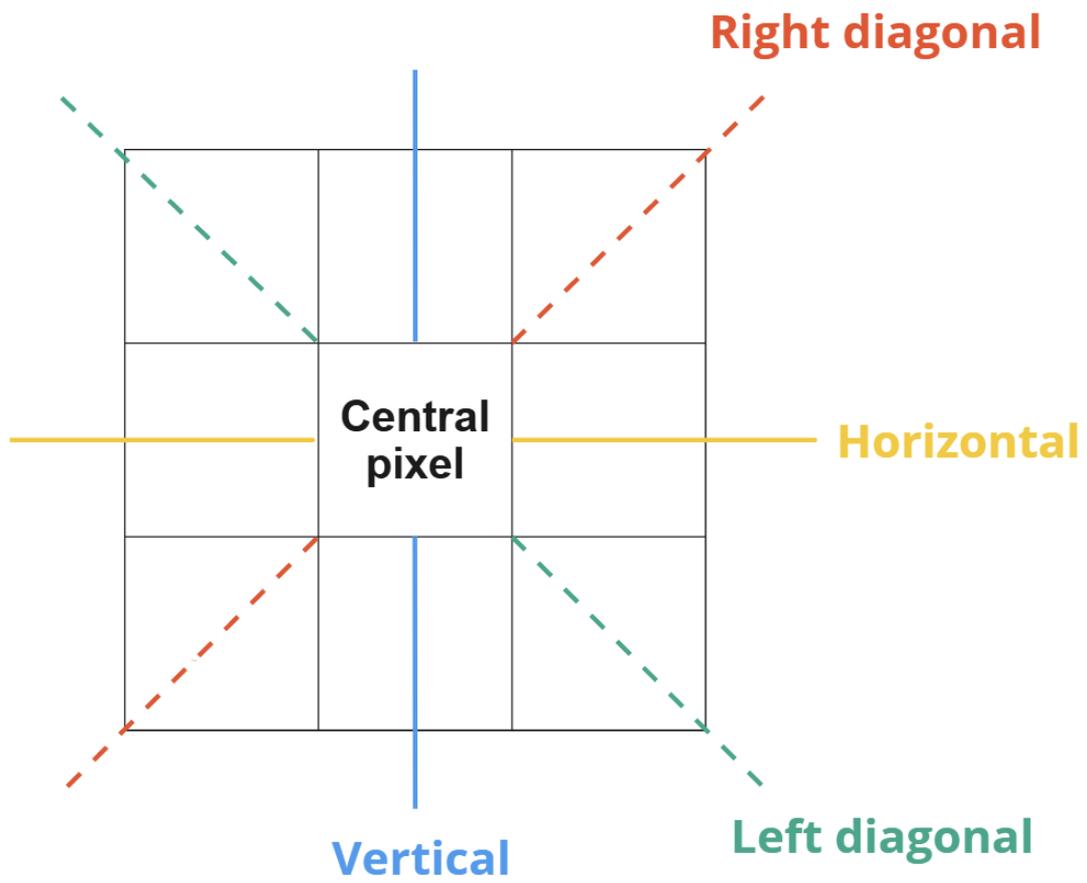

**Figure 4 Directional Pixel Interaction in Feature Extraction for MSCN Coefficients.**

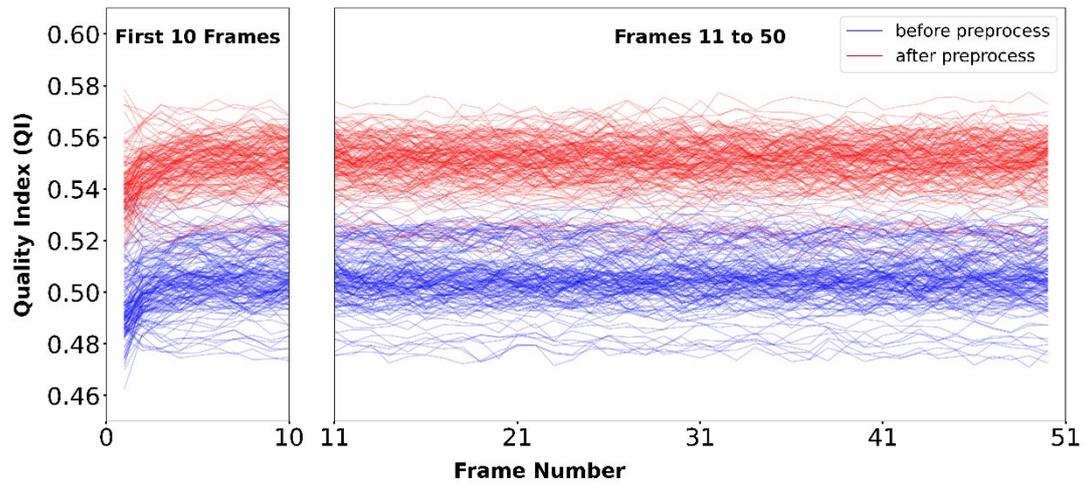

**Figure 5 Comparative Analysis of Image Quality Scores before (Blue Line) and after processing (Red Line).**

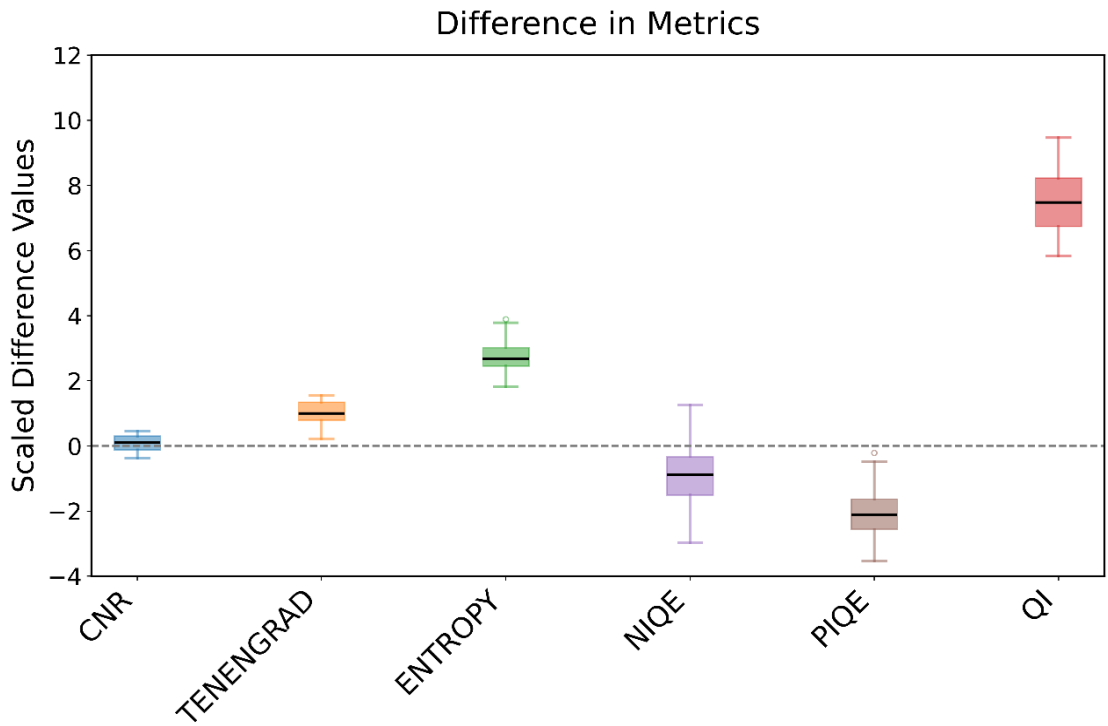

Figure 6 Normalized Difference Values in Various Image Quality Metrics (post-processed minus pre-processed): CNR (Blue), TENENGRAD (Orange), ENTROPY (Green), NIQE (Purple), PIQE (Brown), and QI (Red).